\documentclass{article}

\PassOptionsToPackage{numbers, sort&compress}{natbib}

\usepackage[final]{neurips_2024_ml4ps}

\usepackage[utf8]{inputenc}
\usepackage[T1]{fontenc}
\usepackage{hyperref} 
\usepackage{url} 
\usepackage{booktabs}
\usepackage{amsfonts}
\usepackage{nicefrac} 
\usepackage{microtype} 
\usepackage{xcolor} 
\usepackage{graphicx} 
\usepackage{amsmath}
\usepackage[nameinlink,capitalise]{cleveref}

\title{Gaussian Processes for Probabilistic Estimates of Earthquake Ground Shaking: A 1-D Proof-of-Concept}

\author{
  Sam A.~Scivier \\
  Department of Earth Sciences\\
  University of Oxford\\
  Oxford, UK \\
  \texttt{sam.scivier@earth.ox.ac.uk} \\
  \And
  Tarje Nissen-Meyer \\
  Department of Mathematics and Statistics \\
  University of Exeter \\
  Exeter, UK \\
  \texttt{t.nissen-meyer@exeter.ox.ac.uk} \\
  \AND
  Paula Koelemeijer \\
  Department of Earth Sciences \\
  University of Oxford \\
  Oxford, UK \\
  \texttt{paula.koelemeijer@earth.ox.ac.uk} \\
  \And
  Atılım Güneş Baydin \\
  Department of Computer Science \\
  University of Oxford \\
  Oxford, UK \\
  \texttt{gunes.baydin@cs.ox.ac.uk} \\
}

\begin{document}

\maketitle

\begin{abstract}
  Estimates of seismic wave speeds in the Earth (seismic velocity models) are key input parameters to earthquake simulations for ground motion prediction. Owing to the non-uniqueness of the seismic inverse problem, typically many velocity models exist for any given region. The arbitrary choice of which velocity model to use in earthquake simulations impacts ground motion predictions. However, current hazard analysis methods do not account for this source of uncertainty. We present a proof-of-concept ground motion prediction workflow for incorporating uncertainties arising from inconsistencies between existing seismic velocity models. Our analysis is based on the probabilistic fusion of overlapping seismic velocity models using scalable Gaussian process (GP) regression. Specifically, we fit a GP to two synthetic 1-D velocity profiles simultaneously, and show that the predictive uncertainty accounts for the differences between the models. We subsequently draw velocity model samples from the predictive distribution and estimate peak ground displacement using acoustic wave propagation through the velocity models. The resulting distribution of possible ground motion amplitudes is much wider than would be predicted by simulating shaking using only the two input velocity models. This proof-of-concept illustrates the importance of probabilistic methods for physics-based seismic hazard analysis.
\end{abstract}

\section{Introduction}\label{sec:intro}

Seismic velocity models --- estimates of the Earth's seismic wave speeds --- underpin earthquake ground motion prediction in seismic hazard analysis, as they are key inputs to wave equation solvers. They continue to be produced at different resolutions and scales, stemming from different methods (e.g., tomography \citep{tape2010}, reflection surveys \citep{schultz1984}). The seismic inverse problem is ill-posed as there are not enough data to constrain a unique true Earth model \citep{rawlinson2010}. As such, many overlapping velocity models exist for a given region. Consequently, the choice of which velocity model to use in ground motion prediction is often arbitrary. Nevertheless, it has a significant impact on the results as different models have different structures, length scales, and amplitudes.

The key output of seismic hazard analysis is an estimate of peak ground motion, to assess potential infrastructure damage and inform earthquake engineering. Most commonly, empirical ground motion models (GMMs) \citep{meletti2021, motnikar2022} are used to predict the median and uncertainty of a ground motion parameter (e.g., peak ground displacement --- PGD) for earthquake scenarios \citep{abrahamson2019}. They make rapid predictions, but drastically simplify the underlying physical processes. Importantly, they approximate the effect of seismic velocities on ground motion, typically only using the average shear wave velocity in the uppermost 30~m \citep{graves2010}. GMMs are thus limited in accuracy and reliability. An alternative is to simulate earthquake scenarios in 3-D by solving the wave equation and extracting PGD estimates, requiring 3-D seismic velocity information as input. However, there are two issues: (i) simulating many earthquakes is computationally costly, and (ii) choices of input parameters are subjective, including the input velocity model. To address the first issue, recent advances in machine learning have begun accelerating wave propagation methods \citep{yang2021, moseley2023, ramadan2024, lehmann2024}. However, current physics-based hazard analysis workflows do not consider inconsistencies between velocity models. This omits a key source of uncertainty, given that predicted ground motion can be drastically impacted by velocity structure. One possible solution is to fuse different velocity models, and use the output in earthquake simulations. Unfortunately, existing methods for velocity model fusion \citetext{e.g., \citealp{fichtner2018, ajala2021, zhang2024}} typically do not produce probabilistic outputs, limiting their ability to account for differences between velocity models.

In this study, we propose a workflow to account for inconsistencies between seismic velocity models in ground motion prediction. Our method is based on the probabilistic fusion of velocity models using Gaussian processes (GPs), and estimates uncertainties owing to differences between them. We then produce probabilistic ground motion predictions with respect to these uncertainties by drawing velocity model samples from the GP predictive distribution, simulating acoustic wave propagation using each sample, and extracting the PGD predictions. We illustrate that such a probabilistic method is necessary to capture the spread of possible ground motion scenarios.

Our key contributions are as follows:
(i) We present a workflow for probabilistic earthquake ground motion prediction that accounts for inconsistencies between seismic velocity models.
(ii) We demonstrate the capability of scalable GPs for the probabilistic fusion of different estimates of the same physical parameter, through a synthetic example using 1-D seismic velocity models.
The code for this work is written in Python and is available at \citet{code}.

\begin{figure}[b]
    \centering
    \includegraphics[width=\linewidth]{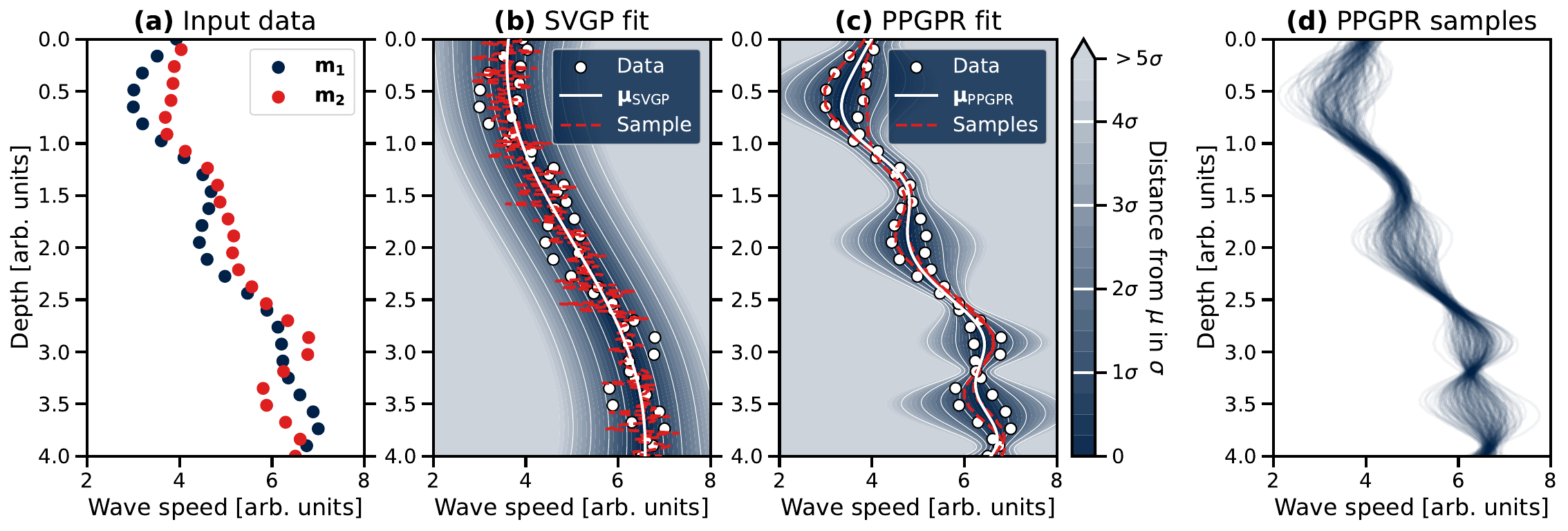}
    \caption{Comparison of SVGPR and PPGPR for the probabilistic fusion of seismic velocity models. (a) shows the input synthetic 1-D seismic velocity profiles with depth. (b) and (c) show the fusion results of SVGPR and PPGPR, and (d) shows the 200 function samples drawn from the PPGPR predictive distribution used in \cref{sec:gmp}. The shading in (b) and (c) show the SVGP posterior predictive distribution, $q_{\text{SVGP}} \left( \mathbf{y_*} \right)$, and the PPGPR latent predictive distribution, $q_{\text{PPGPR}} \left( \mathbf{f_*} \right)$, respectively, in terms of distance from the predictive means in standard deviations.}
    \label{fig:data-fit-samples}
\end{figure}

\section{Gaussian processes and data fusion}\label{sec:methods}

GPs \citep{rasmussen2005} are a class of non-parametric models for defining a distribution over function spaces. They are widely used for regression, providing robust uncertainty quantification and predictive performance. Unfortunately, exact GP regression is limited in scalability owing to a computational cost of $\mathcal{O}(n^3)$, where $n$ is the number of data points. Despite the small datasets used in this study, seismic datasets of realistic size can have $n \sim 10^6\textup{--}10^7$. To overcome this, approximate GP methods have been developed \citep{liu2020}. One popular method for scalable GP inference is the sparse variational Gaussian process (SVGP) \citep{titsias2009, matthews2016}, which applies variational inference to fit GPs. SVGPs introduce a set of inducing variables to approximate the full dataset using a smaller set of points $m \ll n$. Thus, the computational cost is reduced to $\mathcal{O}(nm^2 + m^3)$, making SVGPs practical to apply to large 2-D and 3-D datasets.

In this study, we use scalable GP regression for fusing seismic velocity models by fitting a GP to multiple datasets simultaneously. A key advantage of GPs is their modelling of covariance structure, which enables samples matching the spatial patterns of the input data to be drawn from the predictive distribution.

We aim to model inconsistencies between velocity estimates as uncertainty in the GP predictive distribution. Thus, we need to understand the form of the predictive variance in the SVGP model. Below we provide a brief summary of the key SVGP results to highlight the relevant context for our work. The interested reader is referred to \citet{titsias2009, matthews2016, jankowiak2020}, and \citet{pml2Book} for full derivations and explanations. The inputs are the training set (coordinates of the input velocity models), $\mathbf{X}$, the inducing point locations, $\mathbf{Z}$, and the points at which we wish to predict, $\mathbf{X_*}$. Then $\mathbf{f_X}, \mathbf{f_Z}, \mathbf{f_*}$ are the (unknown) velocity values that we predict at these locations; and $\mathbf{y}$ are the observed data (input velocity values). SVGP-based methods approximate the joint posterior as $q \left( \mathbf{f_*}, \mathbf{f_X}, \mathbf{f_Z} \right) = p \left( \mathbf{f_*}, \mathbf{f_X} \mid \mathbf{f_Z} \right) q \left( \mathbf{f_Z} \right)$, where $p \left( \mathbf{f_*}, \mathbf{f_X} \mid \mathbf{f_Z} \right)$ is calculated exactly \citep{pml2Book}. The variational distribution is $q\left(\mathbf{f_Z}\right) = \mathcal{N} \left( \mathbf{f_Z} \mid \mathbf{m}, \mathbf{S} \right)$, where $\mathbf{m}$\ and $\mathbf{S}$ are (learned) variational parameters. The predictive distribution over the underlying function (at the target points) $\mathbf{f_*}$ is given by
\begin{equation}
    \begin{aligned}
        q \left( \mathbf{f_*} \right) &= \int p\left(\mathbf{f_*} \mid \mathbf{f_Z}\right) q \left( \mathbf{f_Z} \right) \mathrm{d} \mathbf{f_Z} \\
        &= \mathcal{N} \left( \mathbf{f_*} \mid \mathbf{\mu_*}, \mathbf{\sigma_f} \left( \mathbf{x_*} \right) ^2 \right),
    \end{aligned}
\end{equation}

where,

\begin{equation*}
    \begin{aligned}
        \mathbf{\mu_{*}} &= \mathbf{K_{*, Z}} \mathbf{K_{Z, Z}}^{-1} \mathbf{m} \\
        \mathbf{\sigma_f} \left( \mathbf{x_*} \right) ^2 &= \mathbf{K_{*, *}} - \mathbf{K_{*, Z}} \mathbf{K_{Z, Z}}^{-1} \left( \mathbf{K_{Z, Z}} - \mathbf{S} \right) \mathbf{K_{Z, Z}}^{-1} \mathbf{K_{Z, *}},
\end{aligned}
\end{equation*}

with e.g., $\mathbf{K_{*, Z}} = k \left( \mathbf{X_*}, \mathbf{Z} \right)$, and $k \left( \cdot, \cdot \right)$ is the (chosen) covariance function. Assuming a Gaussian likelihood, measurements (at the target points) $\mathbf{y_*}$ are related to the underlying function (at the target points) $\mathbf{f_*}$ as $p \left( \mathbf{y_*} \mid \mathbf{f_*}, \sigma_y^2 \right) = \mathcal{N} \left( \mathbf{y_*} \mid \mathbf{f_*}, \sigma_y^2 \mathbb{I}_* \right)$, where $\sigma_y^2$ is the (learned) observational noise variance. Thus, the predictive distribution over $\mathbf{y_*}$ is,

\begin{equation}
    \begin{aligned}
        q \left( \mathbf{y_*} \right) &= \int p \left( \mathbf{y_*} \mid \mathbf{f_*}, \sigma_y^2 \right) q \left( \mathbf{f_*} \right) \mathrm{d} \mathbf{f_*} \\
        &= \mathcal{N} \left( \mathbf{y_*} \mid \mathbf{\mu_*}, \mathbf{\sigma_f} \left( \mathbf{x_*} \right) ^2 + \sigma_y^2 \mathbb{I}_* \right).
\end{aligned}
\label{eq:pred_post}
\end{equation}

The predictive variance at a target point $\mathbf{x}_{\mathbf{*}, i}$ is thus the sum of input-dependent variance over the underlying function, $\mathbf{\sigma_f} \left( \mathbf{x}_{\mathbf{*}, i} \right) ^2$, and observational noise, $\sigma_y^2$: $\mathrm{Var} \left( \mathbf{x}_{\mathbf{*}, i} \right) = \mathbf{\sigma_f} \left( \mathbf{x}_{\mathbf{*}, i} \right) ^2 + \sigma_y^2$. Despite this symmetry in the predictive variance, \citet{jankowiak2020} highlight that the typical SVGP objective function (variational ELBO) targets only large $\sigma_y^2$ -- often resulting in $\sigma_y^2 \gg \mathbf{\sigma_f} \left( \mathbf{x}_{\mathbf{*}, i} \right) ^2$, which we see from the data-fit term: $\mathcal{L}_{\mathrm{SVGP}} \supset - \frac{1}{2 \sigma_y^2} \left| \mathbf{y}_i - \mu_{\mathbf{X}, i} \right|^2$. This means we would model disagreements between different velocity models as observational noise. Given the degree of disagreement varies spatially, $\sigma_y$ would need to be input-dependent \citetext{e.g., \citealp{liu2020_2, kersting2007}}. However, this would result in noisy samples in the predictive distribution (see \cref{fig:data-fit-samples}b) --- making them useless for downstream tasks. 

We instead wish to model inconsistencies between velocity models as input-dependent uncertainty in the underlying physical process (i.e., $\mathbf{\sigma_f} \left( \mathbf{x} \right)$). To enable this, we use the parametric predictive GP regression (PPGPR) model \citep{jankowiak2020}. PPGPR is a variation of SVGP with an objective function that directly targets the predictive distribution (\cref{eq:pred_post}). Notably, the PPGPR objective encourages large $\mathbf{\sigma_f} \left( \mathbf{x_*} \right) ^2$, as seen from the data-fit term: $\mathcal{L}_{\mathrm{PPGPR}} \supset - \frac{1}{2} \frac{1}{\sigma_y^2 + \mathbf{\sigma_f} \left( \mathbf{x}_{i} \right) ^2} \left| \mathbf{y}_i - \mu_{\mathbf{X}, i} \right|^2$. Contrary to SVGP, this typically results in $\mathbf{\sigma_f} \left( \mathbf{x}_{\mathbf{*}, i} \right) ^2 \gg \sigma_y^2$. Therefore, we can choose to ignore $\sigma_y^2$ and use the PPGPR latent predictive distribution, $q \left( \mathbf{f_*} \right)$.

\section{Fusing synthetic seismic velocity models}\label{sec:merge}

We present a proof-of-concept demonstrating the applicability of PPGPR for the probabilistic fusion of two synthetic 1-D seismic velocity models. We note that we do not consider uncertainties attached to input velocity models in this study (i.e., the input models themselves are not probability distributions). Two datasets, $\mathbf{s}_1$ and $\mathbf{s}_2$, are sampled ($n = 25$ data points, each) from a GP prior, using radial basis function (RBF) kernels with different length scales. The samples have different coordinates in an overlapping region. The first model is set as the first sample, $\mathbf{m}_1 = \mathbf{s}_1$. The second model is a weighted superposition of the two samples, $\mathbf{m}_2 = \frac{2}{3} \mathbf{s}_1 + \frac{1}{3} \mathbf{s}_2$, to create larger-scale similarities and smaller-scale differences --- typical of different seismic velocity models. \cref{fig:data-fit-samples}a shows the input velocity models, which are 1-D profiles with respect to depth.

At training time, the input models are concatenated and used to condition the GP as a single dataset. For regression, we employ both PPGPR and SVGPR for comparison, using scaled RBF kernels, $m = 20$ inducing points (with learned locations), and Gaussian likelihoods. The models are trained using the Adam optimiser \citep{kingma2017} and identical hyperparameters (i.e., learning rate and number of iterations). Hyperparameters are chosen through trial-and-error, and full hyperparameter details are provided in the code \citep{code}. Training the models takes one minute for each method, on a laptop using an NVIDIA T500 2GB GDDR6 GPU.

\cref{fig:data-fit-samples}b and c show the results of SVGPR and PPGPR, respectively, on the two velocity models. As discussed in \cref{sec:methods}, optimising the SVGP objective results in $\sigma_{\mathrm{obs}}^2 \gg \sigma_{\mathbf{f}}(\mathbf{x})^2$, making it unsuitable for this task owing to a lack of input-dependence on the predictive variance and noisy function samples. On the other hand, PPGPR performs well, with predictive samples appearing to reflect the spatial patterns of the input models. In an ideal case of maximum likelihood estimation for fitting a univariate Gaussian distribution to two observations, $y_1$ and $y_2$, the resulting distribution is $\mathcal{N}\left(\mu = \frac{1}{2} (y_1 + y_2), \sigma^2 = \left(\frac{y_1 - y_2}{2}\right)^2\right)$ \citep{pml1Book}. In our example, we therefore expect the $\pm 1\mathbf{\sigma}$ contours to approximately follow each of the input velocity models.
We interpolated $\mathbf{m}_1$ and $\mathbf{m}_2$ at the test points using cubic splines and calculated the root mean square error (RMSE) of the SVGP and PPGPR predictive means and variances with respect to the above ideal result. The RMSEs on $\mathbf{\mu}_{\text{SVGP}}$ and $\mathbf{\sigma}_{\text{SVGP}}^2$ were 0.243 and 0.098 (in wave speed units), respectively, while for $\mathbf{\mu}_{\text{PPGPR}}$ and $\mathbf{\sigma}_{\text{PPGPR}}^2$ the RMSEs were 0.045 and 0.012 (in wave speed units).
The PPGPR predictive distribution thus appropriately quantifies the uncertainty on the knowledge of seismic velocities in the region. Most importantly for our application, the covariance structure of the data is modelled. This enables the drawing of samples from the predictive distribution that match the spatial patterns of the input data. Despite only fusing two velocity models here, our approach is generally applicable for fusing any number of input datasets.

\section{Probabilistic ground motion prediction}\label{sec:gmp}

\begin{figure}
    \centering
    \includegraphics[width=\linewidth]{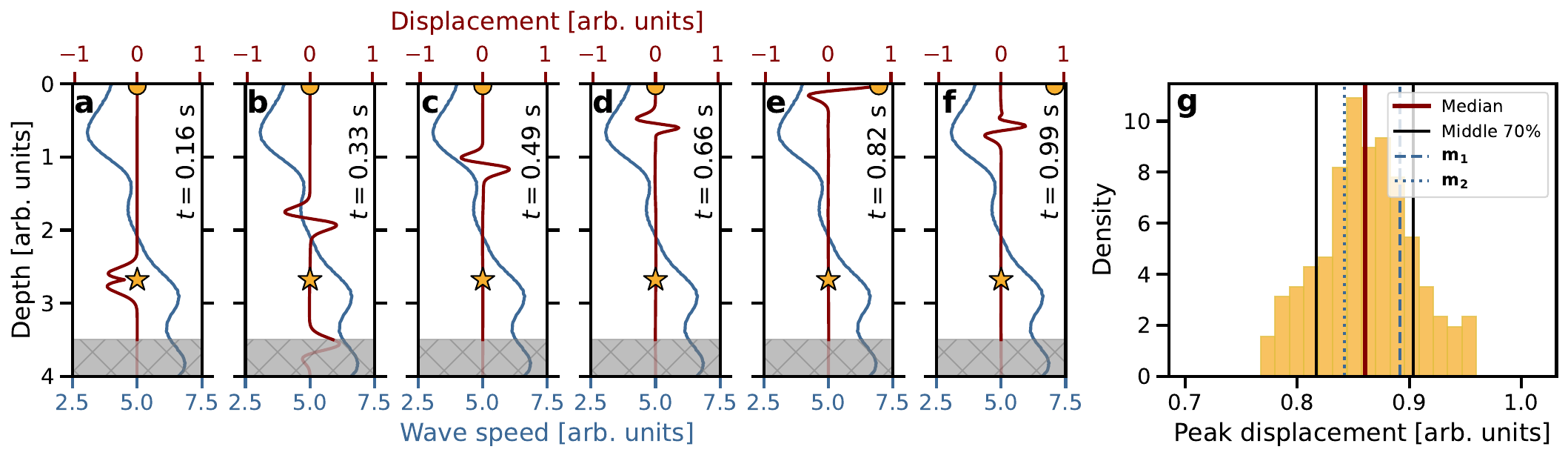}
    \caption{Wavefield snapshots and probabilistic ground motion prediction. (a)--(f) shows wavefield displacement snapshots at increasing time steps for one simulation. Each panel includes the source location (yellow star), the maximum PGD up to that time step (yellow dot), and the underlying velocity model of the simulation. The shaded region indicates where an absorbing boundary layer is applied \citep{chern2019}. (g) shows a histogram of the PGD measurements from the simulations, and highlights the median and middle 70\% of predictions. Also shown are the PGD measurements resulting from simulations using just $\mathbf{m}_1$ or $\mathbf{m}_2$ as input.}
    \label{fig:wavefield-histogram}
\end{figure}

We propose a proof-of-concept workflow for propagating the predicted uncertainty on seismic velocities through simulations of the acoustic wave equation, to produce probabilistic ground motion predictions. First, we draw 200 function samples from the PPGPR latent predictive distribution (\cref{fig:data-fit-samples}d). Then for each sample, we simulate the 1-D acoustic wave equation for displacement, $\mathbf{u}$, with a Ricker wavelet as the earthquake source, using a finite difference scheme (6~s for 200 simulations on a laptop --- vectorised over velocity models).
At the surface, we implement a free-surface boundary condition (i.e., the acoustic pressure $\mathbf{p} = 0$).
At depth, we implement an absorbing boundary layer according to \citet{chern2019}. \cref{fig:wavefield-histogram}a--f shows snapshots of the displacement field at various time steps in one of the simulations. For each simulation, we record the peak ground displacement at the surface (i.e., PGD; depth = 0), producing one PGD estimate per simulation (i.e., per sample velocity model). \cref{fig:wavefield-histogram}g shows a histogram of the recorded PGD measurements from the simulations. Additionally, we ran simulations using interpolated versions of $\mathbf{m}_1$ and $\mathbf{m}_2$ and marked the resulting PGD measurements in \cref{fig:wavefield-histogram}g, to investigate how much information is gained by running simulations for many velocity model samples. Clearly, the PGD measurements for $\mathbf{m}_1$ and $\mathbf{m}_2$ do not account for the spread of possible ground motions, given the degree of knowledge of seismic velocities in this example. Despite being 1-D, our work already shows that it is not possible to approximate the full distribution of possible ground motions using only two velocity models.

\section{Limitations}\label{sec:limitations}

This work is a proof-of-concept and can be extended in several ways. 
For example, we do not account for data with varying length scales or structure, or address kernel design or choice, which would be required for dealing with real seismic datasets. For real-world applicability, it will also be important to extend our workflow from 1-D to 2-D and 3-D, and to solve the elastic wave equation instead of the acoustic  wave equation. Working with 3-D velocity models would add complexity, but in principle it would consist of changing the GP coordinate space from 1-D to 3-D. There are many optimised 3-D seismic wave propagation codes that could then be used for the simulation component of the workflow \citetext{e.g., \citealp{peter2011, maeda2017, leng2019, lehmann2024}}.
If the input velocity models had different spatial densities of data points, the objective function would be weighted towards one of them, and the result would be skewed. This can be readily solved by weighting their contributions to the objective. 
Additionally in this work, the synthetic input velocity models do not have uncertainties, which we plan to incorporate in the future (i.e., the input velocity models would themselves be probability distributions). 

In this study, we were unable to compare our method with existing methods for velocity model fusion. Current methods are generally designed for enhancing larger-scale velocity models using smaller-scale models --- and are thus not applicable when the models occupy the same domain and/or have similar spatial data density, as in our case. Additionally, existing methods typically do not produce probabilistic outputs, meaning it is not possible to compare them in the ground motion prediction component of the study (\cref{sec:gmp}).

\section{Conclusion}\label{sec:conclusion}

Seismic velocity models underpin predictions of earthquake ground motion. Current methods for physics-based seismic hazard analysis do not account for inconsistencies between existing velocity models. In this study, we present a proof-of-concept workflow for probabilistic ground motion prediction that takes this source of uncertainty into account. Firstly, we demonstrate the applicability of scalable GP regression to the probabilistic fusion of input velocity models, showing that inconsistencies between velocity models can be modelled as predictive uncertainty. This provides access to any number of plausible velocity models for the region by drawing samples from the predictive distribution. Secondly, we build up a distribution of possible ground motion scenarios for the family of possible velocity models according to the GP predictive distribution. Our results show a much wider spread of possible peak ground motions than would be predicted by simulating earthquake scenarios using just the input velocity models themselves. We thus highlight the value of using probabilistic methods, such as the one presented here, in physics-based seismic hazard analysis to account for differences between velocity models.

\begin{ack}

\paragraph{Acknowledgments}

The authors thank the five anonymous reviewers for their time and insightful comments, which have improved the quality of the paper. The authors thank Adrian Marin Mag for fruitful discussions and comments on the paper, as well as Fatima Ramadan and Franck Latallerie for useful feedback.

\paragraph{Reproducibility}

The code used to produce the results and figures in this study is available at \citet{code}.

\paragraph{Software}

The code for this work was written in Python, and used the open-source software libaries \textsc{Jupyter Notebooks} \citep{kluyver2016}, \textsc{Binder} \citep{binder2018}, \textsc{NumPy} v2.1.3 \citep{harris2020}, \textsc{SciPy} v1.14.1 \citep{virtanen2020}, \textsc{Matplotlib} v3.9.2 \citep{hunter2007}, \textsc{PyTorch} v2.5.1 \citep{paszke2019}, and \textsc{GPyTorch} v1.13 \citep{gardner2018}.

\paragraph{Funding}

SAS is funded by a UKRI NERC DTP Award (NE/S007474/1) and gratefully acknowledges their support. PK acknowledges financial support from a Royal Society University Research Fellowship (URF\textbackslash R1\textbackslash 180377). AGB was supported by grants from NVIDIA and the Munich Institute for Astro-, Particle and BioPhysics (MIAPbP), funded by the Deutsche Forschungsgemeinschaft under Germany´s Excellence Strategy – EXC-2094 – 390783311.

\paragraph{Competing interests}

The authors have no competing interests to declare.

\end{ack}

\medskip

{
\small

\bibliographystyle{unsrtnat}
\bibliography{refs}

\begin{thebibliography}{36}
\providecommand{\natexlab}[1]{#1}
\providecommand{\url}[1]{\texttt{#1}}
\expandafter\ifx\csname urlstyle\endcsname\relax
  \providecommand{\doi}[1]{doi: #1}\else
  \providecommand{\doi}{doi: \begingroup \urlstyle{rm}\Url}\fi

\bibitem[Tape et~al.(2010)Tape, Liu, Maggi, and Tromp]{tape2010}
Carl Tape, Qinya Liu, Alessia Maggi, and Jeroen Tromp.
\newblock {Seismic tomography of the southern California crust based on
  spectral-element and adjoint methods}.
\newblock \emph{Geophysical Journal International}, 180\penalty0 (1):\penalty0
  433--462, 01 2010.
\newblock ISSN 0956-540X.
\newblock \doi{10.1111/j.1365-246X.2009.04429.x}.
\newblock URL \url{https://doi.org/10.1111/j.1365-246X.2009.04429.x}.

\bibitem[Schultz(1984)]{schultz1984}
P.S. Schultz.
\newblock {Seismic velocity estimation}.
\newblock \emph{Proceedings of the IEEE}, 72\penalty0 (10):\penalty0
  1330--1339, 1984.
\newblock \doi{10.1109/PROC.1984.13021}.

\bibitem[Rawlinson et~al.(2010)Rawlinson, Pozgay, and Fishwick]{rawlinson2010}
N.~Rawlinson, S.~Pozgay, and S.~Fishwick.
\newblock {Seismic tomography: A window into deep Earth}.
\newblock \emph{Physics of the Earth and Planetary Interiors}, 178\penalty0
  (3):\penalty0 101--135, 2010.
\newblock ISSN 0031-9201.
\newblock \doi{https://doi.org/10.1016/j.pepi.2009.10.002}.
\newblock URL
  \url{https://www.sciencedirect.com/science/article/pii/S0031920109002106}.

\bibitem[Meletti et~al.(2021)Meletti, Marzocchi, D'Amico, Lanzano, Luzi,
  Martinelli, Pace, Rovida, Visini, Group, Barreca, Monaco, and
  Iervolino]{meletti2021}
Carlo Meletti, Warner Marzocchi, Vera D'Amico, Giovanni Lanzano, Lucia Luzi,
  Francesco Martinelli, B.~Pace, Andrea Rovida, Francesco Visini, MPS Group,
  Giovanni Barreca, Carmelo Monaco, and Iunio Iervolino.
\newblock {The new Italian Seismic Hazard Model (MPS19)}.
\newblock \emph{{Annals of Geophysics}}, 64\penalty0 (1), 05 2021.
\newblock \doi{10.4401/ag-8579}.

\bibitem[Motnikar et~al.(2022)Motnikar, Zupančič, Živčić, Atanackov,
  Jamšek~Rupnik, Čarman, Danciu, and Gosar]{motnikar2022}
Barbara Motnikar, Polona Zupančič, Mladen Živčić, J.~Atanackov, Petra
  Jamšek~Rupnik, Martina Čarman, Laurentiu Danciu, and Andrej Gosar.
\newblock {The 2021 seismic hazard model for Slovenia (SHMS21): overview and
  results}.
\newblock \emph{Bulletin of Earthquake Engineering}, 20:\penalty0 1--30, 08
  2022.
\newblock \doi{10.1007/s10518-022-01399-8}.

\bibitem[Abrahamson et~al.(2019)Abrahamson, Kuehn, Walling, and
  Landwehr]{abrahamson2019}
Norman~A. Abrahamson, Nicolas~M. Kuehn, Melanie Walling, and Niels Landwehr.
\newblock {Probabilistic Seismic Hazard Analysis in California Using Nonergodic
  Ground‐Motion Models}.
\newblock \emph{Bulletin of the Seismological Society of America}, 109\penalty0
  (4):\penalty0 1235--1249, 07 2019.
\newblock ISSN 0037-1106.
\newblock \doi{10.1785/0120190030}.
\newblock URL \url{https://doi.org/10.1785/0120190030}.

\bibitem[Graves et~al.(2010)Graves, Jordan, Callaghan, Deelman, Field, Juve,
  Kesselman, Maechling, Mehta, Milner, Okaya, Small, and Vahi]{graves2010}
Robert Graves, Thomas Jordan, Scott Callaghan, Ewa Deelman, Edward Field,
  Gideon Juve, Carl Kesselman, Philip Maechling, Gaurang Mehta, Kevin Milner,
  David Okaya, Patrick Small, and Karan Vahi.
\newblock {CyberShake: A Physics-Based Seismic Hazard Model for Southern
  California}.
\newblock \emph{Pure and Applied Geophysics}, 168:\penalty0 367--381, 03 2010.
\newblock \doi{10.1007/s00024-010-0161-6}.

\bibitem[Yang et~al.(2021)Yang, Gao, Castellanos, Ross, Azizzadenesheli, and
  Clayton]{yang2021}
Yan Yang, Angela~F. Gao, Jorge~C. Castellanos, Zachary~E. Ross, Kamyar
  Azizzadenesheli, and Robert~W. Clayton.
\newblock {Seismic wave propagation and inversion with Neural Operators}, 2021.

\bibitem[Moseley et~al.(2023)Moseley, Markham, and Nissen-Meyer]{moseley2023}
Ben Moseley, Andrew Markham, and Tarje Nissen-Meyer.
\newblock {Finite basis physics-informed neural networks (FBPINNs): a scalable
  domain decomposition approach for solving differential equations}.
\newblock \emph{Advances in Computational Mathematics}, 49, 07 2023.
\newblock \doi{10.1007/s10444-023-10065-9}.

\bibitem[{Ramadan} et~al.(2024){Ramadan}, {Fry}, and
  {Nissen-Meyer}]{ramadan2024}
Fatme {Ramadan}, Bill {Fry}, and Tarje {Nissen-Meyer}.
\newblock {Rapid Computation of Physics-Based Ground Motions in the Spectral
  Domain using Neural Networks}.
\newblock In \emph{EGU General Assembly Conference Abstracts}, EGU General
  Assembly Conference Abstracts, page 18444, April 2024.
\newblock \doi{10.5194/egusphere-egu24-18444}.

\bibitem[Lehmann et~al.(2024)Lehmann, Gatti, Bertin, and Clouteau]{lehmann2024}
Fanny Lehmann, Filippo Gatti, Michaël Bertin, and Didier Clouteau.
\newblock {3D elastic wave propagation with a Factorized Fourier Neural
  Operator (F-FNO)}.
\newblock \emph{{Computer Methods in Applied Mechanics and Engineering}},
  420:\penalty0 116718, 2024.
\newblock ISSN 0045-7825.
\newblock \doi{https://doi.org/10.1016/j.cma.2023.116718}.
\newblock URL
  \url{https://www.sciencedirect.com/science/article/pii/S0045782523008411}.

\bibitem[Fichtner et~al.(2018)Fichtner, van Herwaarden, Afanasiev, Simutė,
  Krischer, Çubuk Sabuncu, Taymaz, Colli, Saygin, Villaseñor, Trampert,
  Cupillard, Bunge, and Igel]{fichtner2018}
Andreas Fichtner, Dirk-Philip van Herwaarden, Michael Afanasiev, Saulė
  Simutė, Lion Krischer, Yeşim Çubuk Sabuncu, Tuncay Taymaz, Lorenzo Colli,
  Erdinc Saygin, Antonio Villaseñor, Jeannot Trampert, Paul Cupillard,
  Hans-Peter Bunge, and Heiner Igel.
\newblock {The Collaborative Seismic Earth Model: Generation 1}.
\newblock \emph{Geophysical Research Letters}, 45\penalty0 (9):\penalty0
  4007--4016, 2018.
\newblock \doi{10.1029/2018GL077338}.
\newblock URL
  \url{https://agupubs.onlinelibrary.wiley.com/doi/abs/10.1029/2018GL077338}.

\bibitem[Ajala and Persaud(2021)]{ajala2021}
R.~Ajala and P.~Persaud.
\newblock {Effect of Merging Multiscale Models on Seismic Wavefield Predictions
  Near the Southern San Andreas Fault}.
\newblock \emph{Journal of Geophysical Research: Solid Earth}, 126\penalty0
  (10), 2021.
\newblock \doi{10.1029/2021JB021915}.
\newblock URL
  \url{https://agupubs.onlinelibrary.wiley.com/doi/abs/10.1029/2021JB021915}.

\bibitem[Zhang and Ben-Zion(2024)]{zhang2024}
Hao Zhang and Yehuda Ben-Zion.
\newblock {Enhancing Regional Seismic Velocity Models With Higher-Resolution
  Local Results Using Sparse Dictionary Learning}.
\newblock \emph{Journal of Geophysical Research: Solid Earth}, 129\penalty0
  (1):\penalty0 e2023JB027016, 2024.
\newblock \doi{https://doi.org/10.1029/2023JB027016}.
\newblock URL
  \url{https://agupubs.onlinelibrary.wiley.com/doi/abs/10.1029/2023JB027016}.
\newblock e2023JB027016 2023JB027016.

\bibitem[Scivier et~al.(2024)Scivier, Nissen-Meyer, Koelemeijer, and
  Baydin]{code}
Sam~A. Scivier, Tarje Nissen-Meyer, Paula Koelemeijer, and Atılım~Güneş
  Baydin.
\newblock {Gaussian Processes for Probabilistic Estimates of Earthquake Ground
  Shaking: A 1-D Proof-of-Concept}, November 2024.
\newblock URL \url{https://doi.org/10.5281/zenodo.14246055}.

\bibitem[Rasmussen and Williams(2005)]{rasmussen2005}
Carl~Edward Rasmussen and Christopher K.~I. Williams.
\newblock \emph{{Gaussian Processes for Machine Learning}}.
\newblock The MIT Press, 11 2005.
\newblock ISBN 9780262256834.
\newblock \doi{10.7551/mitpress/3206.001.0001}.
\newblock URL \url{https://doi.org/10.7551/mitpress/3206.001.0001}.

\bibitem[Liu et~al.(2020{\natexlab{a}})Liu, Ong, Shen, and Cai]{liu2020}
Haitao Liu, Yew-Soon Ong, Xiaobo Shen, and Jianfei Cai.
\newblock {When Gaussian Process Meets Big Data: A Review of Scalable GPs}.
\newblock \emph{IEEE Transactions on Neural Networks and Learning Systems},
  31\penalty0 (11):\penalty0 4405--4423, 2020{\natexlab{a}}.
\newblock \doi{10.1109/TNNLS.2019.2957109}.

\bibitem[Titsias(2009)]{titsias2009}
Michalis Titsias.
\newblock {Variational Learning of Inducing Variables in Sparse Gaussian
  Processes}.
\newblock In David van Dyk and Max Welling, editors, \emph{Proceedings of the
  Twelfth International Conference on Artificial Intelligence and Statistics},
  volume~5 of \emph{Proceedings of Machine Learning Research}, pages 567--574,
  Hilton Clearwater Beach Resort, Clearwater Beach, Florida USA, 16--18 Apr
  2009. PMLR.
\newblock URL \url{https://proceedings.mlr.press/v5/titsias09a.html}.

\bibitem[Matthews et~al.(2016)Matthews, Hensman, Turner, and
  Ghahramani]{matthews2016}
Alexander G. de~G. Matthews, James Hensman, Richard Turner, and Zoubin
  Ghahramani.
\newblock {On Sparse Variational Methods and the Kullback-Leibler Divergence
  between Stochastic Processes}.
\newblock In Arthur Gretton and Christian~C. Robert, editors, \emph{Proceedings
  of the 19th International Conference on Artificial Intelligence and
  Statistics}, volume~51 of \emph{Proceedings of Machine Learning Research},
  pages 231--239, Cadiz, Spain, 09--11 May 2016. PMLR.
\newblock URL \url{https://proceedings.mlr.press/v51/matthews16.html}.

\bibitem[Jankowiak et~al.(2020)Jankowiak, Pleiss, and Gardner]{jankowiak2020}
Martin Jankowiak, Geoff Pleiss, and Jacob Gardner.
\newblock Parametric {G}aussian process regressors.
\newblock In Hal~Daumé III and Aarti Singh, editors, \emph{Proceedings of the
  37th International Conference on Machine Learning}, volume 119 of
  \emph{Proceedings of Machine Learning Research}, pages 4702--4712. PMLR,
  13--18 Jul 2020.
\newblock URL \url{https://proceedings.mlr.press/v119/jankowiak20a.html}.

\bibitem[Murphy(2023)]{pml2Book}
Kevin~P. Murphy.
\newblock \emph{Probabilistic Machine Learning: Advanced Topics}.
\newblock MIT Press, 2023.
\newblock URL \url{http://probml.github.io/book2}.

\bibitem[Liu et~al.(2020{\natexlab{b}})Liu, Ong, and Cai]{liu2020_2}
Haitao Liu, Yew-Soon Ong, and Jianfei Cai.
\newblock {Large-scale Heteroscedastic Regression via Gaussian Process},
  2020{\natexlab{b}}.
\newblock URL \url{https://arxiv.org/abs/1811.01179}.

\bibitem[Kersting et~al.(2007)Kersting, Plagemann, Pfaff, and
  Burgard]{kersting2007}
Kristian Kersting, Christian Plagemann, Patrick Pfaff, and Wolfram Burgard.
\newblock {Most likely heteroscedastic Gaussian process regression}.
\newblock In \emph{Proceedings of the 24th International Conference on Machine
  Learning}, ICML '07, page 393–400, New York, NY, USA, 2007. Association for
  Computing Machinery.
\newblock ISBN 9781595937933.
\newblock \doi{10.1145/1273496.1273546}.
\newblock URL \url{https://doi.org/10.1145/1273496.1273546}.

\bibitem[Kingma and Ba(2017)]{kingma2017}
Diederik~P. Kingma and Jimmy Ba.
\newblock {Adam: A Method for Stochastic Optimization}, 2017.
\newblock URL \url{https://arxiv.org/abs/1412.6980}.

\bibitem[{Kevin P. Murphy}(2022)]{pml1Book}
{Kevin P. Murphy}.
\newblock \emph{{Probabilistic Machine Learning: An introduction}}.
\newblock {MIT Press}, 2022.
\newblock URL \url{probml.ai}.

\bibitem[Chern(2019)]{chern2019}
Albert Chern.
\newblock {A reflectionless discrete perfectly matched layer}.
\newblock \emph{Journal of Computational Physics}, 381:\penalty0 91–109,
  March 2019.
\newblock ISSN 0021-9991.
\newblock \doi{10.1016/j.jcp.2018.12.026}.
\newblock URL \url{http://dx.doi.org/10.1016/j.jcp.2018.12.026}.

\bibitem[Peter et~al.(2011)Peter, Komatitsch, Luo, Martin, Le~Goff, Casarotti,
  Le~Loher, Magnoni, Liu, Blitz, Nissen-Meyer, Basini, and Tromp]{peter2011}
Daniel Peter, Dimitri Komatitsch, Yang Luo, Roland Martin, Nicolas Le~Goff,
  Emanuele Casarotti, Pieyre Le~Loher, Federica Magnoni, Qinya Liu, Celine
  Blitz, Tarje Nissen-Meyer, Piero Basini, and Jeroen Tromp.
\newblock {Forward and adjoint simulations of seismic wave propagation on fully
  unstructured hexahedral meshes}.
\newblock \emph{Geophys. J. Int.}, 186\penalty0 (2):\penalty0 721--739, 2011.
\newblock \doi{10.1111/j.1365-246X.2011.05044.x}.

\bibitem[Maeda et~al.(2017)Maeda, Takemura, and Furumura]{maeda2017}
Takuto Maeda, Shunsuke Takemura, and Takashi Furumura.
\newblock {OpenSWPC: an open-source integrated parallel simulation code for
  modeling seismic wave propagation in 3D heterogeneous viscoelastic media}.
\newblock Technical report, Springer, 2017.

\bibitem[Leng et~al.(2019)Leng, Nissen-Meyer, van Driel, Hosseini, and
  Al-Attar]{leng2019}
Kuangdai Leng, Tarje Nissen-Meyer, Martin van Driel, Kasra Hosseini, and David
  Al-Attar.
\newblock {AxiSEM3D: broad-band seismic wavefields in 3-D global earth models
  with undulating discontinuities}.
\newblock \emph{{Geophysical Journal International}}, 217\penalty0
  (3):\penalty0 2125--2146, 02 2019.
\newblock ISSN 0956-540X.
\newblock \doi{10.1093/gji/ggz092}.
\newblock URL \url{https://doi.org/10.1093/gji/ggz092}.

\bibitem[Kluyver et~al.(2016)Kluyver, Ragan-Kelley, P{\'e}rez, Granger,
  Bussonnier, Frederic, Kelley, Hamrick, Grout, Corlay, Ivanov, Avila, Abdalla,
  Willing, and development team]{kluyver2016}
Thomas Kluyver, Benjamin Ragan-Kelley, Fernando P{\'e}rez, Brian Granger,
  Matthias Bussonnier, Jonathan Frederic, Kyle Kelley, Jessica Hamrick, Jason
  Grout, Sylvain Corlay, Paul Ivanov, Dami{\'a}n Avila, Safia Abdalla, Carol
  Willing, and Jupyter development team.
\newblock {Jupyter Notebooks - a publishing format for reproducible
  computational workflows}.
\newblock In Fernando Loizides and Birgit Scmidt, editors, \emph{{Positioning
  and Power in Academic Publishing: Players, Agents and Agendas}}, pages
  87--90. IOS Press, 2016.
\newblock URL \url{https://eprints.soton.ac.uk/403913/}.

\bibitem[{P}roject {J}upyter et~al.(2018){P}roject {J}upyter, {M}atthias
  {B}ussonnier, {J}essica {F}orde, {J}eremy {F}reeman, {B}rian {G}ranger, {T}im
  {H}ead, {C}hris {H}oldgraf, {K}yle {K}elley, {G}ladys {N}alvarte, {A}ndrew
  {O}sheroff, {P}acer, {Y}uvi {P}anda, {F}ernando {P}erez, {B}enjamin~{R}agan
  {K}elley, and {C}arol {W}illing]{binder2018}
{P}roject {J}upyter, {M}atthias {B}ussonnier, {J}essica {F}orde, {J}eremy
  {F}reeman, {B}rian {G}ranger, {T}im {H}ead, {C}hris {H}oldgraf, {K}yle
  {K}elley, {G}ladys {N}alvarte, {A}ndrew {O}sheroff, {M} {P}acer, {Y}uvi
  {P}anda, {F}ernando {P}erez, {B}enjamin~{R}agan {K}elley, and {C}arol
  {W}illing.
\newblock {B}inder 2.0 - {R}eproducible, interactive, sharable environments for
  science at scale.
\newblock In {F}atih {A}kici, {D}avid {L}ippa, {D}illon {N}iederhut, and {M}
  {P}acer, editors, \emph{{P}roceedings of the 17th {P}ython in {S}cience
  {C}onference}, pages 113 -- 120, 2018.
\newblock \doi{10.25080/Majora-4af1f417-011}.

\bibitem[Harris et~al.(2020)Harris, Millman, van~der Walt, Gommers, Virtanen,
  Cournapeau, Wieser, Taylor, Berg, Smith, Kern, Picus, Hoyer, van Kerkwijk,
  Brett, Haldane, del R{\'{i}}o, Wiebe, Peterson, G{\'{e}}rard-Marchant,
  Sheppard, Reddy, Weckesser, Abbasi, Gohlke, and Oliphant]{harris2020}
Charles~R. Harris, K.~Jarrod Millman, St{\'{e}}fan~J. van~der Walt, Ralf
  Gommers, Pauli Virtanen, David Cournapeau, Eric Wieser, Julian Taylor,
  Sebastian Berg, Nathaniel~J. Smith, Robert Kern, Matti Picus, Stephan Hoyer,
  Marten~H. van Kerkwijk, Matthew Brett, Allan Haldane, Jaime~Fern{\'{a}}ndez
  del R{\'{i}}o, Mark Wiebe, Pearu Peterson, Pierre G{\'{e}}rard-Marchant,
  Kevin Sheppard, Tyler Reddy, Warren Weckesser, Hameer Abbasi, Christoph
  Gohlke, and Travis~E. Oliphant.
\newblock {Array programming with {NumPy}}.
\newblock \emph{Nature}, 585\penalty0 (7825):\penalty0 357--362, September
  2020.
\newblock \doi{10.1038/s41586-020-2649-2}.
\newblock URL \url{https://doi.org/10.1038/s41586-020-2649-2}.

\bibitem[Virtanen et~al.(2020)Virtanen, Gommers, Oliphant, Haberland, Reddy,
  Cournapeau, Burovski, Peterson, Weckesser, Bright, {van der Walt}, Brett,
  Wilson, Millman, Mayorov, Nelson, Jones, Kern, Larson, Carey, Polat, Feng,
  Moore, {VanderPlas}, Laxalde, Perktold, Cimrman, Henriksen, Quintero, Harris,
  Archibald, Ribeiro, Pedregosa, {van Mulbregt}, and {SciPy 1.0
  Contributors}]{virtanen2020}
Pauli Virtanen, Ralf Gommers, Travis~E. Oliphant, Matt Haberland, Tyler Reddy,
  David Cournapeau, Evgeni Burovski, Pearu Peterson, Warren Weckesser, Jonathan
  Bright, St{\'e}fan~J. {van der Walt}, Matthew Brett, Joshua Wilson, K.~Jarrod
  Millman, Nikolay Mayorov, Andrew R.~J. Nelson, Eric Jones, Robert Kern, Eric
  Larson, C~J Carey, {\.I}lhan Polat, Yu~Feng, Eric~W. Moore, Jake
  {VanderPlas}, Denis Laxalde, Josef Perktold, Robert Cimrman, Ian Henriksen,
  E.~A. Quintero, Charles~R. Harris, Anne~M. Archibald, Ant{\^o}nio~H. Ribeiro,
  Fabian Pedregosa, Paul {van Mulbregt}, and {SciPy 1.0 Contributors}.
\newblock {{SciPy} 1.0: Fundamental Algorithms for Scientific Computing in
  Python}.
\newblock \emph{Nature Methods}, 17:\penalty0 261--272, 2020.
\newblock \doi{10.1038/s41592-019-0686-2}.

\bibitem[Hunter(2007)]{hunter2007}
J.~D. Hunter.
\newblock {Matplotlib: A 2D graphics environment}.
\newblock \emph{Computing in Science \& Engineering}, 9\penalty0 (3):\penalty0
  90--95, 2007.
\newblock \doi{10.1109/MCSE.2007.55}.

\bibitem[Paszke et~al.(2019)Paszke, Gross, Massa, Lerer, Bradbury, Chanan,
  Killeen, Lin, Gimelshein, Antiga, Desmaison, Köpf, Yang, DeVito, Raison,
  Tejani, Chilamkurthy, Steiner, Fang, Bai, and Chintala]{paszke2019}
Adam Paszke, Sam Gross, Francisco Massa, Adam Lerer, James Bradbury, Gregory
  Chanan, Trevor Killeen, Zeming Lin, Natalia Gimelshein, Luca Antiga, Alban
  Desmaison, Andreas Köpf, Edward Yang, Zach DeVito, Martin Raison, Alykhan
  Tejani, Sasank Chilamkurthy, Benoit Steiner, Lu~Fang, Junjie Bai, and Soumith
  Chintala.
\newblock {PyTorch: An Imperative Style, High-Performance Deep Learning
  Library}, 2019.
\newblock URL \url{https://arxiv.org/abs/1912.01703}.

\bibitem[Gardner et~al.(2018)Gardner, Pleiss, Bindel, Weinberger, and
  Wilson]{gardner2018}
Jacob~R Gardner, Geoff Pleiss, David Bindel, Kilian~Q Weinberger, and
  Andrew~Gordon Wilson.
\newblock {GPyTorch: Blackbox Matrix-Matrix Gaussian Process Inference with GPU
  Acceleration}.
\newblock In \emph{Advances in Neural Information Processing Systems}, 2018.

\end{thebibliography}
}

\end{document}